\documentclass[11pt,colorlinks=true,linkcolor=blue,citecolor=blue, urlcolor=blue]{article}
\usepackage{amsmath,amssymb} 
\usepackage[colorlinks=true,linkcolor=blue!70!black,citecolor=blue!70!black,urlcolor=blue!70!black]{hyperref}

\usepackage{setspace}
\usepackage[T1]{fontenc}
\usepackage[utf8]{inputenc}

\usepackage{booktabs}
\usepackage{doi}
\usepackage[round]{natbib}
\usepackage{caption}
\usepackage{algorithm}
\usepackage{algorithmic}

\usepackage{xcolor}

\usepackage{multirow}
\usepackage{graphicx}
\usepackage{newfloat}
\usepackage{listings}
\DeclareCaptionStyle{ruled}{labelfont=normalfont,labelsep=colon,strut=off}
\floatstyle{ruled}
\newfloat{listing}{tb}{lst}{}
\floatname{listing}{Listing}

\title{\large\bfseries Multi-Platform Aggregated Dataset of Online Communities (MADOC)}

\author{%
    Marija Mitrović Dankulov\textsuperscript{1},\ 
    Aleksandar Tomašević\textsuperscript{3},\ 
    Slobodan Maletić\textsuperscript{2},\\ 
    Miroslav Anđelković\textsuperscript{2},
    Ana Vranić\textsuperscript{1},\ 
    Darja Cvetković\textsuperscript{1},
    Boris Stupovski\textsuperscript{1},\\ 
    Dušan Vudragović\textsuperscript{1},
    Sara Major\textsuperscript{3},\ 
    Aleksandar Bogojević\textsuperscript{1}%
}

\footnotetext[1]{Institute of Physics Belgrade, University of Belgrade, Serbia}
\footnotetext[2]{Vinča Institute of Nuclear Sciences, University of Belgrade, Serbia}
\footnotetext[3]{Faculty of Philosophy, University of Novi Sad, Serbia}

\begin{document}

\maketitle
\vspace{-15pt} 
\begin{abstract}
\small
\noindent The Multi-platform Aggregated Dataset of Online Communities (MADOC) is a comprehensive dataset that facilitates computational social science research by providing a unified, standardized dataset for cross-platform analysis of online social dynamics. MADOC aggregates and standardizes data from four distinct platforms: Bluesky, Koo, Reddit, and Voat, spanning from 2012 to 2024. The dataset includes 18.9 million posts, 236 million comments, and data from 23.1 million unique users across all platforms, with a particular focus on understanding community dynamics and the evolution of toxic behavior across platforms. By providing standardized data structures and FAIR-compliant access through Zenodo, MADOC enables researchers to conduct comparative analyses of user behavior, interaction networks, and content sentiment across diverse social media environments. The unique value of the dataset lies in its cross-platform scope, standardized structure, and rich metadata, making it particularly suitable for studying societal phenomena such as toxic behavior propagation, and user migration patterns in response to platform moderation policies.
\end{abstract}

\newpage

\section{Introduction}

The proliferation of social media platforms has created diverse digital spaces where users interact, share content, and form communities. Understanding these interactions and their societal impact requires comprehensive datasets that span multiple platforms and enable comparative analyses. The Multi-platform Aggregated Dataset of Online Communities (MADOC) addresses this need by providing a standardized collection of user interactions, content, and sentiment data from four distinct platforms: Reddit, Bluesky, Koo, and Voat. Comprising 18.9 million posts and 236 million comments from 23.1 million unique users, MADOC represents one of the largest cross-platform datasets available for social media research. In an era where platform API access has become increasingly restricted—what some researchers call the ``post-API era" \citep{poudel2024naviga-a}—MADOC leverages existing public datasets \citep{Baumgartner_Zannettou_Keegan_Squire_Blackburn_2020,mekacher2022Other,failla2024Other,mekacher2024koo-a} to enable cross-platform research.

The dataset is structured to support various computational social science research directions:

\begin{itemize}
    \item \textbf{Cross-platform User Behavior:} Standardized interaction data enables comparative analysis of user engagement patterns across different platform architectures and community structures \citep{wang2024failed,vranic2023sustai}.
    \item \textbf{Community Dynamics:} Comprehensive community-level data allows researchers to study how communities respond to major events like content moderation, user influx, or platform-wide policy changes, though individual user migration cannot be tracked across platforms due to privacy protections \citep{papasavva2023waitin}.
    \item \textbf{Content and Sentiment Analysis:} Textual content and sentiment scores facilitate research on how discourse and sentiment vary across platforms and communities.
    \item \textbf{Moderation Impact:} Historical data from banned communities provides insight into the effectiveness of platform moderation policies and their impact on user behavior \citep{cima2024great}.
\end{itemize}

MADOC adheres to FAIR principles (Findable, Accessible, Interoperable, and Reusable) through the following key features:

\begin{itemize}
    \item \textbf{Findable:} The dataset is distributed through Zenodo \footnote{\url{https://zenodo.org/records/14637314}} with persistent identifiers and comprehensive metadata for discovery.
    
    \item \textbf{Accessible:} All data and metadata are openly accessible through standard protocols with clear access procedures. We provide Python \footnote{\url{https://pypi.org/project/pymadoc/}}and R \footnote{\url{https://github.com/atomashevic/rMADOC}} packages for seamless access to the dataset.
    
    \item \textbf{Interoperable:} Data is provided in standardized Apache Parquet format with a consistent schema across platforms, enabling seamless integration.
    
    \item \textbf{Reusable:} Detailed documentation of collection methodologies, processing steps, and ethical guidelines supports research reproducibility.
\end{itemize}

MADOC's primary contribution is its methodologically rigorous approach to cross-platform data alignment and standardization. The dataset construction begins with a careful selection of 12 Reddit communities, with corresponding communities later identified on Voat. The first set consists of six general-interest communities (\textit{r/funny, r/gaming, r/pics, r/videos, r/gifs, and r/technology}) that focus on mainstream topics. The second set comprises controversial communities, including four that were banned from Reddit for violating platform policies: \textit{r/fatpeoplehate} (banned in 2015 for systematic harassment), \textit{r/GreatAwakening} (banned in 2018 for inciting violence and promoting conspiracy theories), \textit{r/MillionDollarExtreme} (banned in 2018 for promoting hate speech and alt-right content), and \textit{r/CringeAnarchy} (banned in 2019 following increased hate speech after the Christchurch mosque shootings). Two additional controversial communities, while not banned, have been subjected to increased scrutiny and moderation: \textit{r/KotakuInAction} (heavily moderated since 2014 due to its role in harassment campaigns during Gamergate) and \textit{r/MensRights} (subject to content restrictions due to repeated incidents of coordinated harassment and anti-feminist rhetoric). These communities were chosen specifically because they represent different types of content moderation challenges: from organized harassment campaigns and hate speech to conspiracy theories and extremist content. This selection enables researchers to study both typical online interactions and potentially harmful social dynamics, particularly in communities that have faced different levels of platform enforcement \citep{jhaver2017view}. The selection is strategic: it allows us to analyze how community-level behavior, content patterns, and discourse evolve before and after significant events such as platform bans, content restrictions, or influxes of new users. While individual users cannot be tracked across platforms due to privacy protections and ethical considerations, the dataset captures aggregate changes in community characteristics, sentiment, and content patterns during periods of user migration and community restructuring.


\section{Data Collection and Processing}

\subsection{Data Sources}

MADOC aggregates and standardizes data from four distinct social media platforms: Reddit, Voat, Bluesky, and Koo. 

For Reddit, we utilized the Pushshift.io dataset \citep{Baumgartner_Zannettou_Keegan_Squire_Blackburn_2020}, which encompasses submissions and comments from 2006 to 2020. From this extensive collection, we focused on 12 subreddits: six in the general-interest category, and six controversial.

The Voat dataset \citep{mekacher2022Other}, spanning from 2013 to the platform's shutdown in 2020, provides a complete archive of the platform's content. We extracted data from twelve subverses that directly correspond to our selected Reddit communities, maintaining the same naming convention (e.g., v/funny corresponding to r/funny).

For Bluesky, we incorporated a comprehensive dataset containing over 4 million accounts (81\% of registered users) and their complete posting activity (235M posts) from March 2023 to March 2024 \citep{failla2024Other}. Similarly, we used the existing Koo dataset \citep{mekacher2024koo-a} which encompasses 1.4 million users, 72 million posts, and 75 million comments, spanning from January 2020 to September 2023 . This platform's inclusion is particularly valuable as it represents perspectives from the Global South, with a predominantly Indian user base.

\subsection{Data Processing and Standardization}

A key challenge in creating MADOC was aligning content across platforms with different organizational structures. While Reddit and Voat organize content into topic-specific communities (subreddits and subverses), Bluesky and Koo lack such explicit topic segregation. To address this, we developed a systematic content alignment process using Latent Dirichlet Allocation (LDA) topic modeling \citep{blei2003LDA, maier2018LDA}. The detailed methodology of this process is described in the next section.

The dataset standardization process consists of four main components. First, we implemented comprehensive data cleaning procedures. These include deduplication of entries while preserving temporal precedence, conversion of all date and time related fields to UNIX timestamps, standardization of text encoding and character sets, and cleaning of URLs to remove formatting artifacts. For Bluesky and Koo, we implemented additional URL extraction from content fields, separating embedded links from the main text while preserving both components. To account for the presence of bots on Reddit and Voat—programmed to post, comment automatically, or respond to specific prompts — we implemented a  ``bot removal" procedure. These bots often include the phrase ``I am a bot" in their posts or comments. Therefore, we screened users whose interactions (posts and comments) contained this phrase. If more than 70\% of a user's interactions included ``I am a bot," they were removed from our dataset. The 70\% threshold was chosen after analyzing the number of users flagged as bots across various detection thresholds (ranging from 40\% to 100\%) for all communities. We observed a sharp decrease in flagged users at 70\% for in every community, indicating it as the most suitable threshold for identifying bots across all communities.
 
Second, we standardized interaction types across platforms. All interactions were classified as either posts, comments, or reposts (the latter only for Bluesky and Koo). Posts are submissions on Reddit and Voat and tweet-like posts on Koo and Bluesky. Comments include submission comments on Reddit and Voat and replies to posts on Koo and Bluesky. 

For Voat comments, we had only parent information linking posts and first-level comments (excluding comments on comments). This is because the original Voat dataset does not contain complete information about comment-to-comment relationships. To maintain a consistent network structure across platforms, we preserved only first-level replies and comments, excluding nested conversations. This standardization required careful validation of parent-child relationships in threaded discussions, particularly for platforms like Reddit and Voat where deep conversation trees are common. We kept comments that had no parent information, because they relate to posts which have been deleted or banned from the platform.

The absence of parent information is noticeable on Reddit where, due to increased content-moderation, bans, and user migrations, original post-to-comment relationships are not always preserved. The most extreme case is r/fatpeoplehate subreddit, where all posts are missing because the content was banned.

Third, we processed all textual content through multiple stages. We applied VADER sentiment analysis \citep{hutto2014vader} to quantify the emotional valence of each interaction, following its successful application in social media analysis. For the initial release, we restricted the dataset to English-language content. For Koo, which serves a predominantly multilingual user base \citep{mekacher2024koo-a}, we implemented additional language filtering steps to ensure reliable sentiment analysis results.

Fourth, we implemented privacy protection measures. All post and user identifiers across all platforms are pseudonymized using UUID-based hashing, which maintains referential integrity while preventing casual re-identification. The only personally identifiable information is the content of the posts along with the timestamp of the post. However, this information was already present in the original datasets, so our pseudonymization makes it harder to reconstruct the original content in comparison with the original datasets.

This approach maintains the network structure and enables research on user behavior patterns while protecting user privacy through pseudonymization. 

The resulting dataset follows a standardized schema across all platforms, with the structure detailed in Table~\ref{tab:schema}. 

\begin{table}[htbp]
\centering
\begin{tabular}{@{}lp{8cm}@{}}
\toprule
Field & Description \\
\midrule
post\_id & Unique identifier for the interaction, anonymized to prevent reconstruction of original URLs \\
\hline
publish\_date & UNIX timestamp of the interaction (seconds since epoch) \\
\hline
user\_id & Anonymized identifier of the content creator, consistent across all interactions from the same user \\
\hline
parent\_id & Identifier of the parent post for comments/replies, NA for original posts \\
\hline
parent\_user\_id & Identifier of the parent post's creator, NA for original posts \\
\hline
content & Textual content of the interaction, with URLs extracted \\
\hline
url & External URLs referenced in the content\\
\hline
language & Language of the content \\
\hline
interaction\_type & Type of interaction: `POST', `COMMENT', or `REPOST' \\
\hline
platform & Source platform: `reddit', `voat', `bluesky', or `koo' \\
\hline
community & Community identifier (subreddit/subverse name), NA for Bluesky/Koo \\
\hline
sentiment\_vader & VADER sentiment score ranging from -1 (negative) to 1 (positive) \\
\hline
strict\_filter & Whether content matches strict keyword filtering criteria (TRUE/FALSE) \\
\hline
\end{tabular}
\caption{Structure of the MADOC dataset. Each row represents a single interaction (post, comment, or repost) with the fields standardized across all platforms.}
\label{tab:schema}
\end{table}

\section{Topic analysis and selection of posts}

To enable cross-platform content analysis, we developed a systematic approach for identifying thematically similar content across platforms with different organizational structures. While Reddit and Voat organize content into topic-specific communities (subreddits and subverses), Koo and Bluesky lack explicit topic segregation. We employed Latent Dirichlet Allocation (LDA) topic modeling \citep{blei2003LDA, maier2018LDA} to extract characteristic keywords from Reddit and Voat communities, which were then used to identify related content on Koo and Bluesky.

For each community pair (subreddit and corresponding subverse), we created a combined corpus by merging posts and their first-level comments from both platforms. Each document in the corpus represents a post with its comments, preprocessed by removing mentions, hyperlinks, non-alphanumeric characters, converting to lowercase, removing stopwords, and lemmatizing. We excluded words appearing in more than 99\% of documents following \citep{maier2018LDA} and filtered-out documents with fewer than 20 or more than 2000 words to ensure LDA performance \citep{tang14LDA}.

Due to platform size disparities, we implemented a balanced sampling strategy based on document volume:
\begin{itemize}
    \item For communities with similar document counts ($<$10,000): Created nine balanced samples of 20,000 documents (10,000 from each platform)
    \item For communities with 10,000-20,000 Voat documents: Created five samples matching Voat's volume with Reddit documents
    \item For communities with $<$10,000 Voat documents: Created five samples of 20,000 Reddit documents plus all Voat documents
\end{itemize}

We determined the optimal number of topics for each community by maximizing coherence scores, then extracted the top 20 words per topic across all samples. The resulting keyword lists were filtered to retain only the least frequent third of English words \citep{norvig2009natura} to improve topic specificity.

For content alignment, we applied two filtering criteria to Koo and Bluesky posts:
\begin{itemize}
    \item Basic filter: Documents containing at least one community keyword
    \item Strict filter: Documents containing at least two different community keywords
\end{itemize}


The topic modeling analysis revealed distinct linguistic patterns characteristic of each community, providing strong validation for our cross-platform content alignment approach. For general-interest communities, the extracted keywords closely matched their intended focuses. For example, r/gaming exhibited gaming-specific terminology (e.g., ``pokemon'', ``hack''), while r/technology contained science and technology terms (e.g., ``solar'', ``nasa'', ``electricity'').

The topic modeling also revealed distinct linguistic patterns in communities focused on specific social discussions. For instance, r/MensRights showed a concentration of legal and family-related terminology (``father'', ``parent'', ``judge''). Communities focused on political discourse showed distinctive patterns. The r/GreatAwakening community, which centered on conspiracy theories before its ban in 2018, featured high frequencies of governance-related terms (``congress'', ``campaign'', ``election'', ``truth''). Similarly, r/KotakuInAction, which emerged during the GamerGate controversy \citep{jhaver2017view} and focused on media criticism, developed its own specialized vocabulary around content moderation and journalistic practices. These distinctive vocabulary patterns proved particularly valuable for identifying thematically similar content on platforms lacking explicit community structures.

The topic modeling analysis also revealed patterns of problematic language usage across platforms. We observed systematic variations in the frequency and context of inflammatory rhetoric, derogatory terms, and exclusionary language. These patterns were particularly pronounced in communities that were eventually subject to platform moderation actions. This shows that the sampling strategy we used is effective in recognizing the specific language employed by communities on Voat.

These results demonstrate the robustness of our topic modeling and keyword filtering methodology for cross-platform content alignment. The approach successfully captures both explicit topical focus (through technical and domain-specific terminology) and implicit community characteristics (through platform-specific linguistic patterns), enabling meaningful comparative analyses across diverse social media environments.

\section{Dataset Statistics} \label{sec:stats}

MADOC provides a comprehensive view of user interactions across four distinct platforms, with data spanning multiple years and interaction types. Table~\ref{tab:platform_stats} presents the key statistics for each platform, including temporal coverage, interaction counts, and user activity metrics.

\begin{table}[ht!]
\small
\centering
\setlength{\tabcolsep}{6pt}
\begin{tabular}{|l|r|r|r|r|}
\hline
\textbf{Metric} & \textbf{Reddit} & \textbf{Voat} & \textbf{Bluesky} & \textbf{Koo} \\
\hline
Time span & 2014-2020 & 2013-2020 & 2023-2024 & 2020-2023 \\
\hline
Total interactions & 247.6M & 1.2M & 2.8M & 4.3M \\
\hline
Posts & 14.5M & 0.4M & 0.9M & 3.1M \\
\hline
Comments & 233.1M & 0.8M & 0.9M & 1.2M \\
\hline
Unique users & 22.6M & 0.1M & 0.2M & 0.2M \\
\hline
Avg. posts per user & 2.7 & 8.5 & 14.4 & 23.0 \\
\hline
Avg. comments per user & 11.6 & 7.8 & 6.2 & 10.6 \\
\hline
Mean sentiment & 0.063 & 0.011 & 0.088 & 0.054 \\
\hline
\% with URLs & 10.3\% & 31.8\% & 4.5\% & 11.6\% \\
\hline
\end{tabular}
\caption{Platform-level statistics for MADOC. The dataset covers different time periods for each platform, reflecting their operational histories and data availability.}
\label{tab:platform_stats}
\end{table}

The platform-level statistics reveal significant differences in scale and user behavior across platforms. Reddit dominates in terms of total interactions (247.6M) and user base (22.6M), being orders of magnitude larger than the other platforms. However, smaller platforms show higher per-user engagement rates, with Koo users averaging 23.0 posts per user compared to Reddit's 2.7. Voat shows notably higher URL sharing (31.8\% of posts) compared to other platforms (4.5-11.6\%), suggesting a stronger focus on external content sharing. Sentiment analysis reveals that Bluesky has the most positive average sentiment (0.088), while Voat shows the lowest (0.011), potentially reflecting differences in community norms and content moderation approaches.

For Reddit and Voat, where content is organized into topic-specific communities, we provide detailed statistics for each community pair in Table~\ref{tab:community_stats}. These statistics enable direct comparisons between equivalent communities across the two platforms.

\begin{table}[H]
\small
\centering
\resizebox{\columnwidth}{!}{%
\begin{tabular}{|l|r|r|r|r|r|r|r|r|}
\hline
\multirow{2}{*}{\textbf{Community}} & \multicolumn{4}{c|}{\textbf{Reddit}} & \multicolumn{4}{c|}{\textbf{Voat}} \\
\cline{2-9}
& \textbf{Posts} & \textbf{Comments} & \textbf{Users} & \textbf{Avg. Sent.} & \textbf{Posts} & \textbf{Comments} & \textbf{Users} & \textbf{Avg. Sent.} \\
\hline
funny & 4.2M & 57.6M & 5.5M & 0.056 & 47K & 97K & 20K & 0.001 \\
\hline
gaming & 2.9M & 41.4M & 4.1M & 0.114 & 29K & 41K & 12K & 0.093 \\
\hline
pics & 2.4M & 49.5M & 4.9M & 0.075 & 15K & 24K & 9K & 0.037 \\
\hline
videos & 3.1M & 33.7M & 3.2M & 0.054 & 45K & 41K & 14K & -0.037 \\
\hline
gifs & 449K & 20.6M & 2.8M & 0.051 & 8K & 14K & 6K & 0.029 \\
\hline
technology & 848K & 10.9M & 1.3M & 0.055 & 35K & 49K & 19K & 0.029 \\
\hline
fatpeoplehate & 0 & 1.4M & 79K & 0.011 & 75K & 279K & 22K & 0.002 \\
\hline
GreatAwakening & 67K & 848K & 26K & 0.079 & 103K & 222K & 12K & 0.016 \\
\hline
MillionDollarExtreme & 69K & 1.2M & 28K & 0.008 & 7K & 14K & 3K & 0.000 \\
\hline
CringeAnarchy & 195K & 6.3M & 342K & -0.031 & 1K & 2K & 1K & -0.038 \\
\hline
KotakuInAction & 128K & 6.7M & 146K & -0.029 & 3K & 5K & 2K & -0.028 \\
\hline
MensRights & 148K & 3.0M & 185K & -0.076 & 2K & 2K & 1K & -0.162 \\
\hline
\end{tabular}
}
\caption{Community-level statistics for Reddit and Voat. The first six rows show general-interest communities, while the last six show controversial communities. Average sentiment scores range from -1 (negative) to 1 (positive).}
\label{tab:community_stats}
\end{table}

The community-level comparison between Reddit and Voat reveals several interesting patterns. First, there is a clear scale difference, with Reddit communities typically being 100-1000 times larger than their Voat counterparts. For instance, r/funny has 5.5M users compared to v/funny's 20K. Second, sentiment patterns show that general-interest communities (e.g., gaming, pics) tend to maintain positive sentiment scores on both platforms, with Reddit generally showing more positive values. In contrast, controversial communities exhibit negative sentiment scores across both platforms, with some communities like MensRights showing particularly negative values (Reddit: -0.076, Voat: -0.162). The absence of posts in Reddit's r/fatpeoplehate while having 1.4M comments reflects the community's ban during the data collection period, providing an interesting case study of community dynamics around platform moderation actions.

For Bluesky and Koo, which lack explicit community structures, we provide statistics based on the keyword filtering approach described in previous section. Using the basic filtering criteria (one keyword match), we identified 910,376 posts and 932,545 comments from Bluesky, and 3,083,191 posts and 1,180,449 comments from Koo that align with the topics of the Reddit/Voat communities. With the strict filtering criteria (two keyword matches), these numbers reduce to 114,431 posts and 322,193 comments for Bluesky, and 249,983 posts and 138,418 comments for Koo, respectively.

To facilitate efficient data access and analysis, we store the dataset in Apache Parquet format, with separate files for each community (subreddit/subverse) while Koo and Bluesky are stored in a single file. This modular organization enables researchers to easily load specific subsets of the data while maintaining the ability to perform cross-platform analyses. The standardized schema ensures that files can be seamlessly combined (concatenated) across platforms and communities, allowing flexible dataset construction based on research needs. For example, researchers can combine all controversial communities across platforms, merge specific platform pairs for comparative analysis, or create custom subsets based on temporal or content criteria. The standardized structure also allows for the construction of interaction networks where nodes represent users and edges represent their interactions through posts and replies, enabling comparative analysis of community structures across platforms.

\section{Ethical Considerations}

The MADOC dataset raises several important ethical considerations that we have addressed throughout its development and release. Our approach follows the guidelines for responsible data science and research ethics \citep{gebru2021datash}, with particular attention to privacy protection, content handling, and potential misuse prevention.

\textbf{Data Availability:} The dataset is publicly available at \url{https://zenodo.org/records/14637314}.

The dataset is created through aggregation of existing, publicly available datasets, and we have taken extensive measures to enhance privacy protections beyond those present in the source data. We have not introduced any new sources of personally identifiable information (PII). Instead, all user identifiers are pseudonymized using UUID-based hashing, which maintains referential integrity while preventing casual re-identification. Platform-specific identifiers have been standardized and pseudonymized to prevent cross-platform user tracking. The only retained potentially identifying information is post content and timestamps, which were already public in the source datasets.

The aggregation of data complies with the existing licenses of previously published datasets, where licensing information is available (Reddit Pushshift under CC0; Voat, Bluesky, and Koo under CC BY 4.0 International). The MADOC dataset is released under CC BY 4.0 International License to ensure broad accessibility while maintaining attribution requirements.

The dataset includes offensive content present in post and comment text, URLs attached to posts, and community names and descriptions, which may be harmful to some users. However, the inclusion of this content is necessary for several important research purposes: enabling cross-platform studies of toxic and hateful content propagation, understanding the effectiveness of different moderation approaches, studying community dynamics around controversial topics, and analyzing patterns of harmful behavior across platforms. This content is particularly valuable for researchers studying the evolution and spread of harmful narratives across different social media environments.

We acknowledge several potential risks of misuse of this dataset, including: use of the data to train harmful AI models or content generation systems, analysis aimed at identifying vulnerable communities or users, attempts to re-identify users across platforms, and exploitation of toxic content patterns for harassment. To mitigate these risks, we have implemented several safeguards. These include clear documentation of appropriate use cases and research guidelines, strict pseudonymization protocols that make re-identification significantly more difficult, release through established academic repository (Zenodo) with usage tracking, and detailed metadata and documentation following FAIR principles.

Despite these risks, we believe the dataset provides significant value for legitimate research purposes. It enables understanding of cross-platform content moderation effectiveness, studying community migration patterns and their impact, developing better detection systems for harmful content, and improving platform design to promote healthier online interactions. We encourage researchers using MADOC to carefully consider these ethical implications in their work and to implement appropriate safeguards in their research designs. Following standard ethical guidelines and making no attempt to re-identify users, we have focused on creating a resource that balances research utility with responsible data stewardship.

\section{Limitations}

There are several limitations of our dataset production approach. First, the choice of Reddit communities, especially the non-controversial ones, was arbitrary and based on the 20 most active general-interest communities which existed on both Reddit and Voat. In the next versions of the dataset, we aim to include more Reddit-Voat community pairs.

Second, the LDA topic modelling approach was not cross-validated by performing topic modelling again on the filtered communities to confirm the match between the communities. Furthermore, we have not systematically tested and evaluated different possible solutions of the topic modelling approach, apart from selecting for the best coherence score. Future versions of the dataset will provide a more robust and systematic approach to topic modelling for cross-platform content alignment.

Third, we only introduce VADER sentiment scores in this version of the dataset. Future versions will provide a more comprehensive overview of the posts' sentiment using different approaches (TextBlob, RoBERTa-based models, etc.).

Finally, while we have aggregated a substantial amount of data, there may be gaps in our collection due to the limitations of archival sources and API restrictions. Some content might have been deleted or modified before collection, and we cannot guarantee complete coverage of all interactions that occurred on the platform.

\section{Conclusion}

In this work we present MADOC, a comprehensive cross-platform dataset comprising 18.9 million posts and 236 million comments from 23.1 million unique users across four distinct social media platforms: Reddit, Voat, Bluesky, and Koo. The dataset spans from 2013 to 2024 (matching platform operational periods) and represents one of the largest standardized collections of cross-platform social media interactions. We combine data from multiple sources and implement standardization procedures to ensure consistency and comparability across platforms.

The dataset's unique value lies in its standardized structure and rich metadata, which facilitate various types of computational social science research. The inclusion of both mainstream and alternative platforms, along with comprehensive historical data on banned communities, makes MADOC particularly valuable for studying platform governance, content moderation, and user migration patterns. Our careful curation process and adherence to FAIR principles ensure the dataset's accessibility and reusability while maintaining privacy protections through pseudonymization protocols.

We envision MADOC enabling several important research directions: First, researchers can use it to study how content moderation policies affect user behavior and community dynamics across different platform architectures. Second, the dataset's temporal breadth supports longitudinal analyses of how online discourse and community norms evolve over time and across platforms. Third, the standardized sentiment scores and content features facilitate comparative studies of how platform design influences user interactions and content toxicity.

Beyond these immediate applications, we believe that MADOC will contribute to a broader understanding of online social phenomena. The dataset can help answer questions about community formation, user activity patterns, and the propagation of both harmful and beneficial narratives across platform boundaries. For computational researchers, it provides a rich testbed for developing and evaluating natural language processing models, particularly in areas like toxic content detection, cross-platform user behavior prediction, and community dynamics modeling.

Through this work, we aim to support both quantitative and qualitative research into online social dynamics. We encourage researchers to use MADOC responsibly, considering the ethical implications discussed in this paper, and to build upon our methodology for future cross-platform dataset development. To facilitate easy adoption, we provide Python (pyMADOC) and R (rMADOC) packages with command-line interfaces that enable selective downloading of data by platform or community. As social media continues to evolve and fragment across multiple platforms, we believe standardized cross-platform datasets like MADOC will become increasingly valuable for understanding and improving online social spaces.

\section{Acknowledgments}

This research was financially supported by the Science Fund of Republic of Serbia, Prizma programme (grant No. 7416).

Data processing was performed on PARADOX-IV supercomputing facility  at the Scientific Computing Laboratory, National Center of Excellence for the Study of Complex Systems, Institute of Physics Belgrade.


\bibliography{aaai25}
\bibliographystyle{plainnat}

\end{document}